\documentclass[useAMS]{mn2e} 
\usepackage{graphicx}

\newcommand{\aap}{A\&A}  
  
\newcommand{\aj}{AJ}  
\newcommand{\apj}{ApJ}  
  
\newcommand{\apjs}{ApJS}

\newcommand{\mnras}{MNRAS}

\def \kms{\ifmmode{~{\rm km\,s}^{-1}}\else{~km~s$^{-1}$}\fi}  
\def \vhel{\ifmmode{V_{{\rm hel}}}\else{$V_{{\rm hel}}$}\fi}  
\def \vsys{\ifmmode{V_{{\rm sys}}}\else{$V_{{\rm sys}}$}\fi}  
\def \vlsr{\ifmmode{V_{{\rm lsr}}}\else{$V_{{\rm lsr}}$}\fi}  
\def \vobs{\ifmmode{V_{{\rm obs}}}\else{$V_{{\rm obs}}$}\fi}  
\def \degree{\ifmmode{^{\circ}}\else{$^{\circ}$}\fi}  
\def \lsun{\ifmmode{{\rm\ L}_\odot}\else{${\rm\ L}_\odot $}\fi}  
\def \msun{\ifmmode{{\rm\ M}_\odot}\else{${\rm\ M}_\odot$}\fi}  
\def \myr{\ifmmode{{\rm\ M}_\odot{\rm\ yr}^{-1}}\else{${\rm\ M}_\odot$   
yr$^{-1}$}\fi}  
\def \teff{\ifmmode{{\rm{T}}_{\rm eff}}\else{${\rm{T}}_{\rm eff}$}\fi}  
\def \mdot{\ifmmode{{\rm\dot{M}}}\else{${\rm\dot{M}}$}\fi}

\newcommand{\ha}{H$\alpha$}

\newcommand{\HNII}{{\rm H}$\alpha+$[N {\sc ii}]~6548~\&~6584~\AA}

\newcommand{\nii}{[N~{\sc ii}]~6584~\AA}

\newcommand{\NII}{[N~{\sc ii}]~6548~\&~6584~\AA}
\newcommand{\SII}{[S~{\sc ii}]~6716~\&~6731~\AA} 
  
\newcommand{\oiii}{[O~{\sc iii}]~5007~\AA}  
\newcommand{\sii}{[S~{\sc ii}]~6717 \& 6731~\AA}

\title[]{Towards an explanation for the 30 Dor (LMC) Honeycomb nebula --
the impact of recent observations and spectral analysis.}  
\author[J. Meaburn et al]{J. Meaburn$^{1}$\thanks{E-mail:
jmeaburn@jb.man.ac.uk}, M. P. Redman$^{2}$,  P. Boumis$^{3}$
and E. Harvey$^{2}$\\
$^{1}$Jodrell Bank Centre for Astrophysics, University of Manchester,  
Manchester M13 9PL, UK.\\ 
$^{2}$Centre for Astronomy,School of Physics, National 
University of Ireland Galway,University Road, Galway, Ireland.\\
$^{3}$Institute of Astronomy \& Astrophysics, National Observatory of  
Athens, I. Metaxa \& V. Pavlou, GR--152 36 P. Penteli, Athens,  
Greece.\\  
}  
  
\begin{document}  
  
\date{Received; Accepted}  
  
\pagerange{\pageref{firstpage}--\pageref{lastpage}} \pubyear{2010}  
  
\maketitle  
  
\label{firstpage}  
  
\begin{abstract}  
{The unique Honeycomb nebula, most likely a peculiar supernova
remnant, lies in 30 Doradus in the Large Magellanic Cloud. Due to its
proximity to SN1987A, it has been serendipitously and intentionally
observed at many wavelengths.  Here, an optical spectral analysis of
forbidden line ratios is performed in order to compare the Honeycomb
high--speed gas with supernova remnants in the Galaxy and the LMC,
with galactic Wolf--Rayet nebulae and with the optical line emission from the
interaction zone of the SS433 microquasar and W50 supernova remnant
system.  An empirical spatiokinematic model of the images and spectra
for the Honeycomb reveals that its striking appearance 
is most likely due to a
fortuitous viewing angle. The Honeycomb nebula is more extended in
soft X-ray emission and could in fact be a small part of the edge of a
giant LMC shell revealed for the first time in this short wavelength
domain. It is also suggested that a previously unnoticed region of
optical emission may in fact be an extension of the Honeycomb around
the edge of this giant shell. A secondary supernova explosion in the
edge of a giant shell is considered for the creation of the Honeycomb
nebula. A microquasar origin of the Honeycomb nebula as opposed to a
simple supernova origin is also evaluated.}
\end{abstract}  
\section{Introduction}  
The 30 Doradus nebula in the Large Magellanic Cloud (LMC) is the most
massive and largest H {\sc ii} complex in the Local Group of
galaxies. Its 300 pc diameter core and adjacent halo exists between
two 1000 pc diameter supergiant filamentary shells (Meaburn 1979;
1980). The multitude of young massive stars scattered throughout this
region identify it as a nursary of recent star formation consequently,
the supernova (SN) rate is of particular interest.The explosion of 
 SN1987A emphasised
that these events are ongoing.

The SN rate in this nebular complex has been considered by several
authors recently. In particular Lazendic, Dickel \& Jones (2003) using
high resolution optical, radio and X-ray imagery searched for young
supernova remnants (SNRs) in its central 80 pc diameter region but
came up with only four regions with high radio/\ha\ ratios.
 However, two of these were associated with young HII regions with
young embedded stellar objects. Chu et al. (2004) then showed the remaining
two SNR candidates to be dust clouds or obscured star forming regions.
This leaves N157
along with the Honeycomb nebula (Wang 1992) as the only firm SNR candidates 
but  many more young (1000 yr
old) remnants are likely to be present but as yet undetected. This
probability lead Meaburn (1984; 1988) to suggest that such remnants
could be identified by the high--speed (200-300 \kms) velocity
`spikes', with predominantly approaching radial velocities, that are
present in the longslit position--velocity (pv) arrays of optical line
profiles (e.g. see fig.3 of Meaburn 1988). Chu et al. (1994) suggested
quite correctly that these are not necessarily diagnostic of SN
activity because wind--blown shells, particularly around Wolf--Rayet
(WR) stars can have similar extents in radial velocity. Maybe
individual SNRs just become merged and combine with particle winds to
drive the giant (50--100 pc diameter) shells found around the many
OB-associations in the 300 pc diameter halo of 30 Doradus.

However, it seemed worthwhile to investigate if optical line
brightness ratios of only the high--speed component of the 30 Doradus
radial velocity spikes could be used as a diagnostic of individual SN
origin. This opportunity became viable for it was realised that in
previous, separate observations of the Honeycomb nebula (Wang 1992),
itself most likely of SN origin for it is a non--thermal radio source
(Chu et al. 1995 though see Sect. 4.4 for a more radical alternative
possibility), that one slit position for \ha\ and \NII\ longslit line
profiles (Meaburn et al. 1993) matched, within the `seeing' disk, the
position for the \sii\ profiles (Redman et al. 1999). Sound
diagnostic optical line ratios of the high--speed ionized gas
therefore became accessible.

In the present paper these line ratios of the Honeycomb nebula's
high--speed gas are evaluated and they resemble those of LMC but not
Galactic SNRs. The creation and structure of the strange, and
possibly unique, Honeycomb nebula itself, is re-considered also aided
by the most recent Chandra and other X--ray imagery
plus Hubble Space Telescope (HST) and ESO New Technology Telescope (NTT)
imagery. The impact of
these results on an evaluation of the 30 Doradus SNe rate is also
considered.
\section{Observations}
Due to it's location in the vicinity of SN1987A, the Honeycomb nebula
has been observed by XMM-Newton and Chandra in the X-ray regime and in
narrow band optical imaging with the 
NTT. Figure 1(a) shows narrow band images of the Honeycomb in \ha\
plus \NII. Figure 1(b) is an \oiii\ image, in which the Honeycomb is
seen more extensively. Figure 1(c) is a Chandra X-ray image of the
source in which it can be seen that the spatial extent of the X-ray
emission closely matches the optical emission. Part of the Honeycomb
was also serendipitously observed at the very edge of the field by the Hubble
Space Telescope (HST) as
part of a project to detect light echos from SN1987A (Crotts
1988). These broad--band observations are of much lower 
sensitivity than the narrow-band ones in Figs. 1a--b but
do show that the edges of the
Honeycomb cells are not resolved even at the resolution of the
HST. The edge widths to cell diameter ratios are remarkably $\leq$ 0.01.
(see Sect.4.2 and Fig. 7).

\ha\ and \NII\ (Meaburn et al. 1993) and \sii\ (Redman et al. 1999)
line profiles were obtained using the Manchester Echelle Spectrometer
(MES; Meaburn et al. 1984) on the Anglo-Australian telescope along the
same (within the one arcsec wide seeing disk) slit position
(Fig. 2a). The position--velocity (pv) array of \sii\ profiles along
this slit is shown in Fig. 2b. The atmospheric conditions were
photometric in both cases. The details of these observations and their
analyses are fully described in the respective papers and will not be
repeated here.
\section{Optical line brightness ratios}
The two MES long slit pv arrays of line profiles had been converted
into absolute surface brightnesses ($B~{\rm erg~s^{-1}~cm^{-2} sr^{-1}
\AA^{-1}}$) using the spectra of standard stars. The brightnesses
along the slit length of each of the five emission lines were obtained
in the heliocentric radial velocity ranges of \vhel\ $=$ 150--230
\kms\ and 240--300 \kms\ using YSTRACT in the STARLINK FIGARO suite of
data reduction programmes. The example pv array in Fig. 2b reveals
that the former velocity range contains only the high--speed Honeycomb
features whereas the latter range contains the emission from
predominantly the 30 Doradus host nebula. The peaks of the $B$ values for
all five lines in the 150--230 \kms\ range occurred at the positions
marked 1-6 in Fig. 2b and each is coincident with a high--speed
velocity `spike' from the Honeycomb nebula.  The log$_{10}$ [\ha\ /
(\NII) ] versus log$_{10}$ [\ha\ /(\SII)] brightness ratios for the
high speed `spikes' are shown in Fig. 3 for positions 1--6 as square
dots. Similarly the \SII\ ratios I($\lambda$ 6717)/I($\lambda$ 6731)
versus log$_{10}$ [\ha\ /(\NII)] and log$_{10}$[\ha\ /(\SII)] ratios
are shown in Figs 4 and 5 respectively. The mean ratios for the 30
Dor host nebula from the same data are marked in each of Figs 3--5 by
a cross. All of these values are compared with the regions occupied by
line ratios of Galactic SNRs, H {\sc ii} regions and planetary nebulae
(PNe) as given in the diagnostic diagrams of Sabbadin, Minello \&
Bianchini (1977). The values for LMC SNRs (Payne, White \& Filipovic
2008) are shown as diamonds. The large open circles and large black
dots only in Fig. 3 are for the Galactic filamentary WR nebulae NGC
3199 and RCW 104 as given by Whitehead, Meaburn \& Goudis (1988) and
Goudis, Meaburn \& Whitehead (1988) respectively. One black dot for
RCW 104 is of particular interest for it is for a high speed
knot. This is shown in Fig. 3 with an arrow attached to show its
limiting ratio as \sii\ was not detected.

The uncertainties in the Honeycomb log$_{10}$ [\ha\ /(\NII)],
log$_{10}$[\ha\ /(\SII)] and I($\lambda$ 6717)/I($\lambda$ 6731)
brightness ratios are $\pm$ 0.04, $\pm$ 0.04 and $\pm$ 0.05
respectively.  Those for the brighter emission from the host 30 Doradus
nebula are significantly smaller.
\section{Discussion}
\subsection{Line ratio diagnostics}
The Honeycomb nebula is a non--thermal radio source and therefore most
likely of SN origin (Chu et al. 1994; 1995) though see Sect. 4.4 for a
more radical, but unlikely, possibility. However, it is striking that
the positions of the line ratios of the Honeycomb high--speed gas, on
the diagnostic diagrams in Figs. 3--5 match most closely those
occupied by LMC but not Galactic SNRs. These positions are
significantly away from those for Galactic SNRs before considering
that the nitrogen abundance of the LMC  is lower by a factor of 2
compared with that of the Galaxy (Russell \& Dopita 1992). This could
proportionally lower the \nii\ brightness from the Honeycomb
nebulosity and move line ratios closer to the Galactic SNR zones in
Figs. 3 \& 5.

Furthermore, ratios for the high speed Honeycomb gas bear no
resemblance to those of the two WR nebulae considered here (Fig. 3)
even when Galactic and LMC abundance differences are taken into
account. Even the high--speed component of the Galactic WR nebula, RCW
104, is far removed from the Honeycomb zone. When determining the SN
rate for the halo of the 30 Doradus nebula from the number of high--speed
velocity spikes over the region (Meaburn 1984; 1988) any confusion
with those of WR origin is easily clarified by the relative positions
of their line ratios on diagnostic diagrams.
\subsection{Modelling kinematics and morphology}
The SHAPE code of Steffen (see Steffen, Holloway \& Pedlar 1996 for
its initial use
and Steffen \& L{\'o}pez 2006) permits the actual structure 
and kinematics of a
nebula to be deduced from the imagery (Fig. 1 a--b) and long--slit pv
arrays (such as that shown in Fig. 2b). No consideration of emission
mechanisms or of the dynamics is involved here but the spatiokinematic
modelling is useful to investigate how critical the viewing angle is
to the unique Honeycomb morphology.

The Honeycomb was modelled as a set of cylinders with equal length and
diameter (since the Rayleigh--Taylor instability, for example, tends to
grow most rapidly for those modes of order of the thickness of the
disturbed layer). The edges of some of the Honeycomb cells are visible
in archive HST images yet are unresolved. The edge widths to cell
diameter ratios are $\leq$ 0.01.

The Chandra images (eg. Fig. 1c) are of comparable resolution to the
size of the cells but do seem more indicative of emission from the
centre of each cell rather than poorly resolved boundary layers. A
simple model would then be of hot X--ray emitting gas venting through
gaps in a preexisting shell or layer of gas and accelerating the
prexisting material in a boundary layer at the cell edges. Therefore,
the Honeycomb cells were modelled with SHAPE as a series of very thin,
nested optically emitting cylinders, with each cylinder inwards having
an increasing characteristic velocity. The blueshifted velocity spikes
(eg. Fig. 2b) then result from this range of velocities being
simultaneously present within the unresolved edges of the Honeycomb
cells. While the Honeycomb morphology and kinematics can be reproduced
in such an empirical model, it should be stressed that it does not
describe the radiation emission or hydrodynamics of the system.

More interesting perhaps is to examine what happens when the model
system described above is viewed from a different angle. Figs. 6a--c
show that the circular cells, when viewed from increasing viewing
angle, rapidly overlap and become blended into arcs and knots,
somewhat reminiscent of the tangled filamentary structure that is
commonly seen in optical emission lines toward the edges of
SNRs. Similarly, the striking velocity spikes characteristic of the
Honeycomb blend into more confused and diffuse emission on the pv
array. If the material forming the cell edges is being continually
entrained and eroded then the extreme thinness of the cell edges
suggests that the uniqueness of the Honeycomb could simply be due to
it being a short--lived structure viewed from a fortuitous angle.
 The low cell edge width to diameter ratio of the most prominent
cell is clear in the HST image in Fig. 7 and should be compared to
its synthetic image in Fig. 6a. 

However, the Honeycomb may not be being viewed exactly in the plane of
the sky. The individual cells are almost all distinctly brighter on
their western sides which is easily reproduced by the SHAPE code if
they are viewed at a small angle to the line of sight. Furthermore,
there are redshifted velocity spikes present in the Honeycomb that are
not explained by the empirical model above though see possible
explanation in Sect. 4.3.
\subsection{A SN explosion in an expanding LMC giant shell}
It was thought that the Honeycomb appearance and the pv arrays across
the nebula could only be produced as a secondary SN blast wave
encountered clumps of material in the nearside of a preceding
expanding shell (e.g. see fig. 3 of Redman et al. 1999). This
interpretation can certainly explain the nearly circular features in
the imagery and the predominance of approaching velocity `spikes' in
the pv arrays of line profiles which are coincident with the edges of
the sub--shells in the nebular image.

However, it fails to explain the elongated, three ridge structure that
is particularly evident in the Chandra X--ray image in Fig. 1c and the
fact that positive velocity spikes occur over the most westerly ridge
of Honeycomb shells which are most apparent in the \oiii\ image in
Fig. 1b. Another possibility therefore is that such a secondary SN
explosion has occurred in the edge of a preceding giant LMC 
shell but now viewed
tangentially to this previous feature (see Fig. 8). The three ridges
of secondary Honeycomb shells could then be in folds of the surface of
the original shell. Direct, but partial, evidence for the existence of
this preceding $\approx$ 8\arcmin\ ($\equiv$ 130 pc) diameter giant
shell centred on RA 05h 36m 30s DEC $-$69\degree\ 19\arcmin\ (J 2000)
can be seen in the X-ray imagery in fig. 5 of Dennerl et al. (2001),
fig 1{\it l} of Dunne, Points \& Chu (2001) and fig. 2 of Smith
\& Wang (2004). The Honeycomb nebula forms a small part of its
western edge but only a northern ridge is clearly detected at X--ray
wavelengths. The X-ray image of Dennerl et al. (2001) shows a bright
feature at the Honeycomb location that also clearly extends in an arc
to the northeast. This extended X--ray feature is compared in Fig. 9
to the area of the images in Figs. 1a--c. Strangely, fainter
Honeycomb--like features are also present in optical images along
parts of this ridge mostly outside the area of Figs 1a--c
(but see top-left of Fig.1a)
 but as yet no line
profiles have been obtained from them to confirm their Honeycomb
nature.
\subsection{A microquasar origin}
The unusual, and possibly unique, morphology and kinematics of the
Honeycomb nebula warrants the consideration of a more radical
possibility than given in Sect. 4.3 for its origin. Could it be the
manifestation of the collision of a precessing relativistic jet
emitted by a binary microquasar similar to SS 433 (for a review see
Fabrika 2004), Cygnus X--1 (Russel et al. 2007), IC 342 X--1 (Feng \&
Kaaret 2008) and that in the galaxy NGC 7793 (Soria et al. 2009)?
Non--thermal radio emission as seen emanating from the Honeycomb
nebula would also be expected in this interaction zone between the jet
and the ambient gas.

The eight hard X-ray point sources found by Haberl et al. (2001) in
the vicinity of the Honeycomb nebula, could be candidates for the
origin of such a jet. Although Haberl et al. (2001) strongly favour background
active galactic nuclei for the origin of these point, hard X--ray
sources, they do not completely rule out the microquasar possibility
in all cases. 

 The faintest of these eight point sources 
(Source 1 of Haberl et al. 2001) 
has no measured spectral index and can be seen in their fig. 5
to be well placed with respect to the Honeycomb nebula if such a
microquasar mechanism is occurring and if this nebula 
is on the western edge of an elongated structure with this source central.
A much fainter candidate is the marginally detected point source arrowed in 
Fig. 9 but seen more clearly in fig. 5 of Haberl et al. (2001) though not
listed by these authors. This is towards the centre of the more spherical
giant shell proposed here to have the Honeycomb nebula at its edge.

 However, it is interesting that
the optical line ratios for the Galactic SS 433/W50 nebulosity (Boumis
et al. 2007) when placed into the diagnostic diagrams shown in
Figs. 3--5 bear no resemblance to the Honeycomb ratios or any other
phenomena except Galactic planetary nebulae. Perhaps, the X--rays from
the SS 433 jet are radiatively ionizing the processed material of the
W50 SNR envelope. Furthermore, the different  X-ray properties of the 
Honeycomb Nebula 
and W50 do not support a microquasar origin for the former. 
Even though these points argue against a microquasar origin
for the Honeycomb nebula it will still be worthwhile to see if any 
stars within the error boxes
of the
two point X--ray sources in Fig. 9 have  microquasar characteristics.
\section{Conclusions}
The optical line ratios of the high--speed Honeycomb nebula confirm
its most likely SNR origin.

\noindent 
The velocity spikes on pv arrays of longslit line profiles found over
the rest of the halo of 30 Doradus can be identified as of young SNR
origin if they also occupy the LMC SNR zone on such diagrams. In this
way they can be distinguished from high--speed WR shells.

\noindent
The circular structures of the Honeycomb gas and their corresponding
approaching and receding, highly collimated, flows can be modelled as
a young SNR in the edge of a larger and preceding giant shell. The
appearance and kinematics of the separate Honeycomb cells is strongly
dependent on viewing angle i.e.they are only apparent if the cell
walls are viewed along their cylindrical axes.

\noindent
An alternative view that this unique nebula could be the consequence
of a microquasar jet (similar to that from SS 433) is shown from
comparative optical line ratios, and by other arguments, 
to be highly  unlikely but not completely
dismissed as there are two  point X--ray sources in its vicinity, and
because it is such a strange object.

\noindent
The positions of the line ratios of the SS 433/W 50 interaction 
nebulosity on the
diagnostic diagrams are remarkable in their own right.

. 
\section*{Acknowledgments}
We thank Wolfgang Steffen for help using the SHAPE coded and Ravi
Sankrit for past useful discussions and advice on Chandra data.
  
% 

%  
  
% 

%-------------------------------------------------------------------------  
 
\newpage  
  
%-----------------------------------------------------
\begin{figure*}
\centering
\scalebox{0.75}{\includegraphics{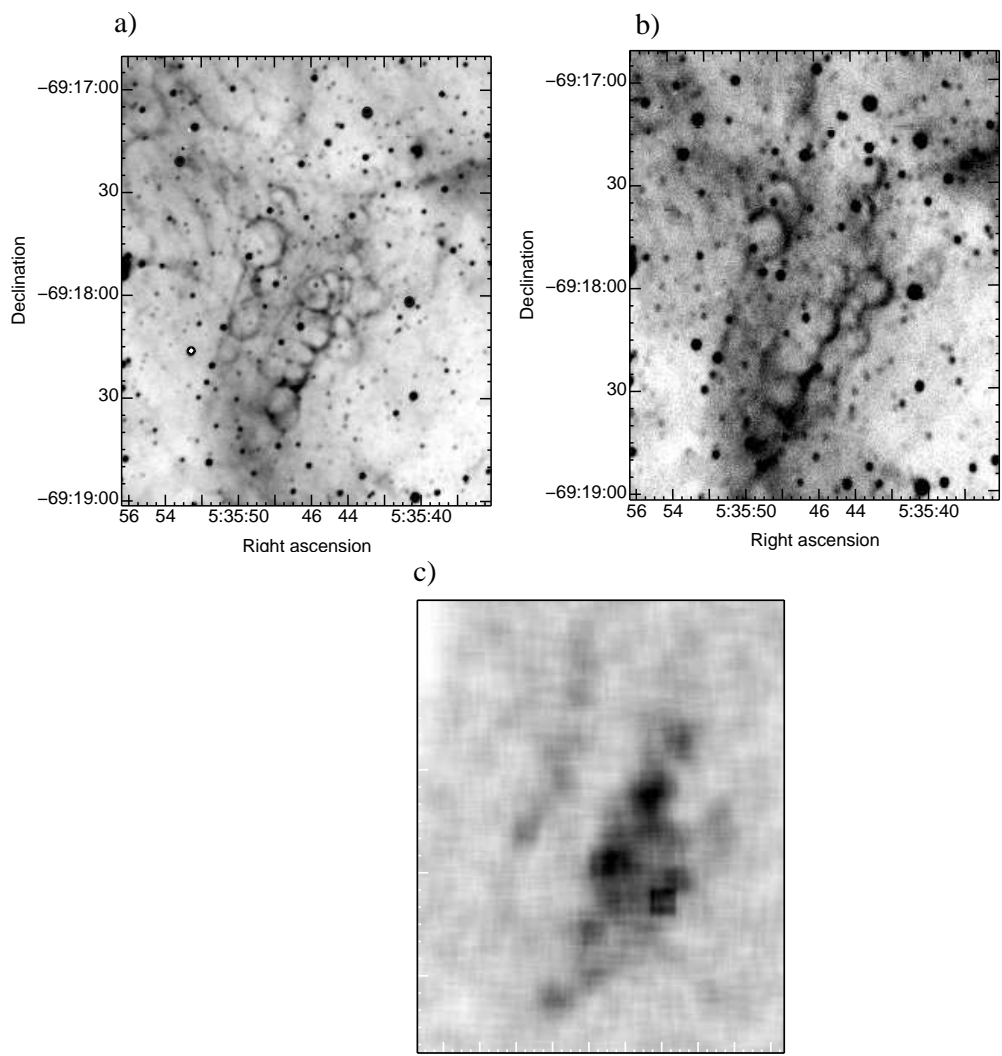}}  
\caption[]{The a) \HNII\ and b) \oiii\ images of the Honeycomb nebula
are compared with c) the Chandra X--ray image. All are for the 
same area of the sky. (J2000 coords).}
\label{fig1}
\end{figure*}
%------------------------------------------------------

\begin{figure*}
\centering
%\scalebox{0.75}{\includegraphics{Fig2.eps}}  
\scalebox{0.75}{\includegraphics{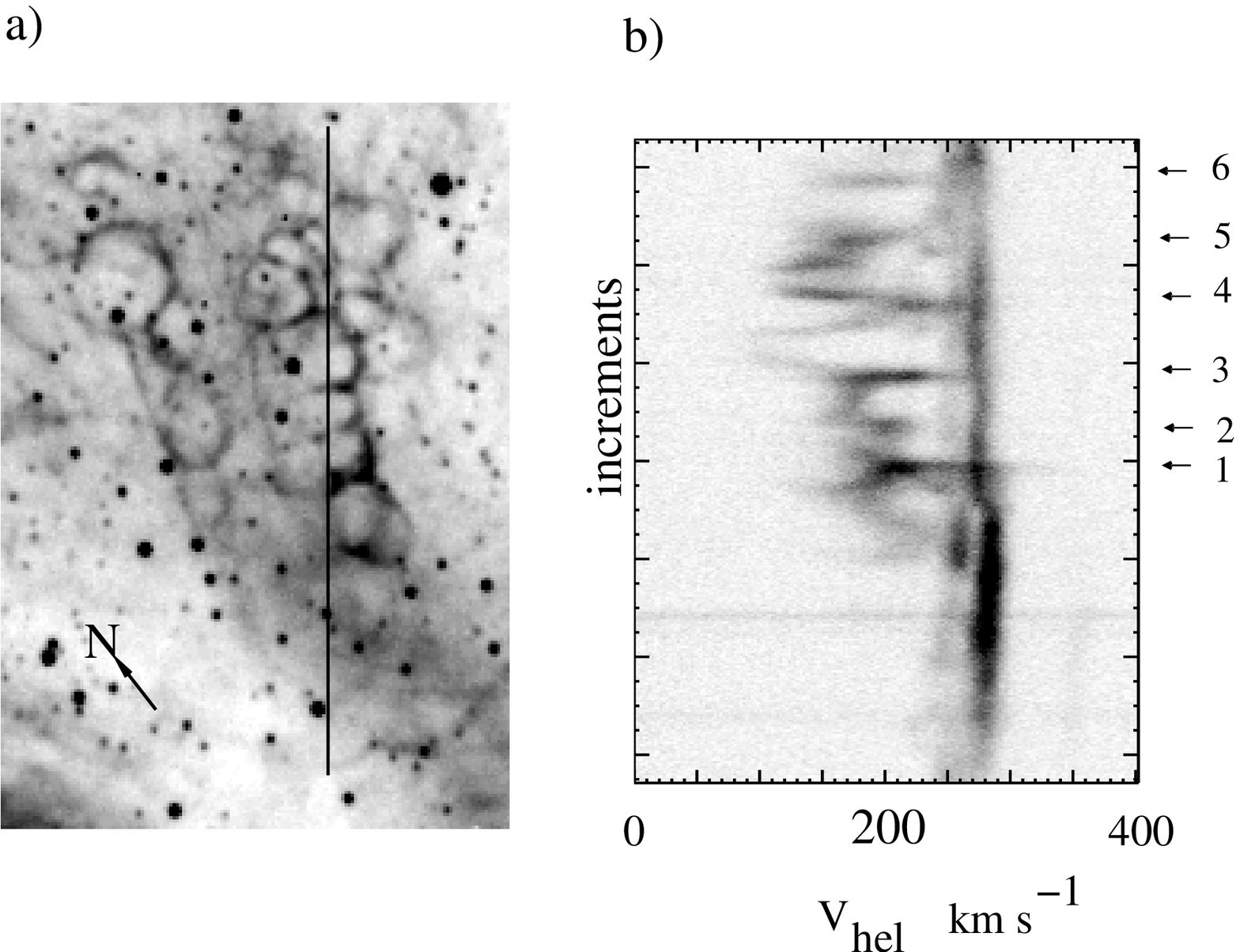}}  
\caption[]{a) The position of part of the longslit is shown against
the \HNII\ image of the Honeycomb nebula and compared in b) with the
pv array of \sii\ profiles along this same length.The positions 1--6
coincident with negative velocity `spikes' where line ratios were
measured are also indicated.}
\label{fig2} 
\end{figure*}
%--------------------------------------------------
\begin{figure*}
\centering
\scalebox{0.7}{\includegraphics{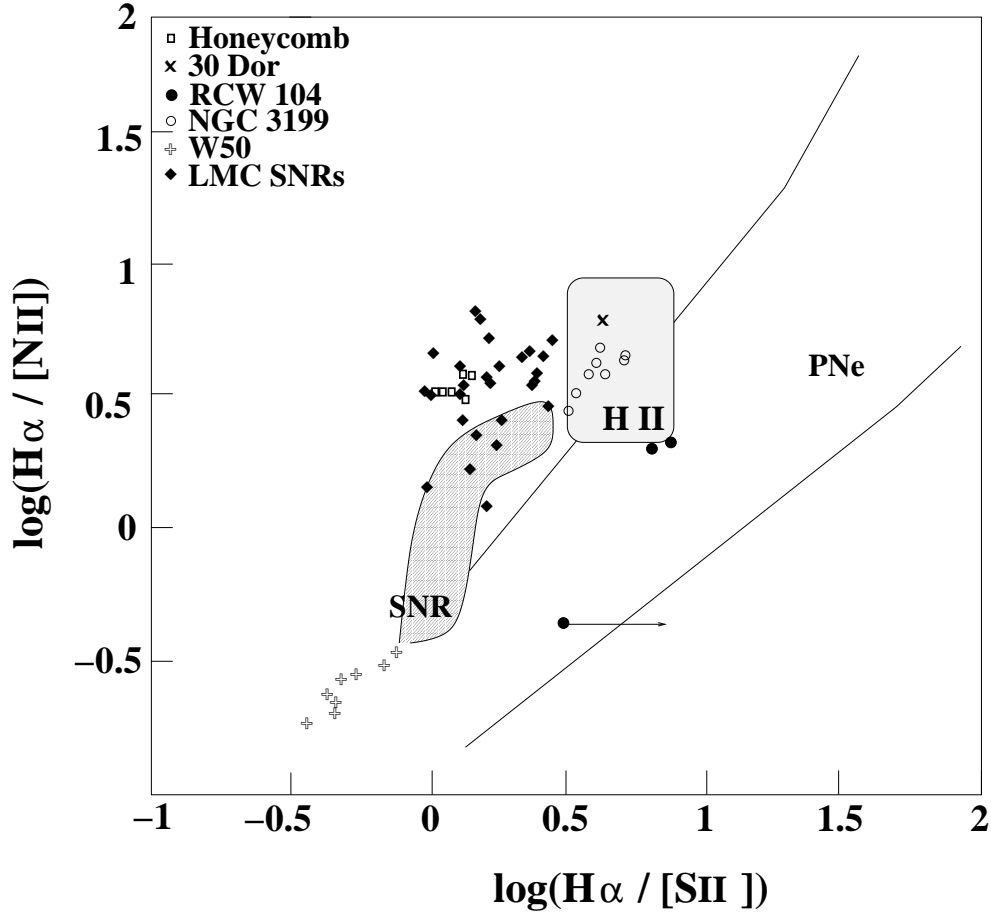}}
%\scalebox{0.7}{\includegraphics{Fig3NF.eps}}  
%\scalebox{0.7}{\includegraphics{Fig3NF.eps}}  
\caption[]{The six line intensity ratios (square dots) measured for
 positions 1--6 (Fig. 2b) for the high speed components in the
 Honeycomb line profiles are compared with the areas of this
 diagnostic diagram occupied by galactic SNRs, H {\sc ii} regions and
 planetary nebulae. The mean of the line ratios for 30 Doradus in the same
 vicinity is marked as a cross. Also shown are line ratios for two
 galactic Wolf--Rayet nebulae RCW 104 and NGC 3199. The large dot with
 an arrow is for the high--speed gas in RCW 104 and the arrow
 indicates that the \sii\ lines were not detected. All other WR values
 are for the filamentary features. The values for LMC SNRs are shown
 as diamonds. The ratios for the W 50 nebulosity generated by the
 microquasar, SS 433, are shown by plus signs.}
\label{fig3} 
\end{figure*}
%-------------------------------------------------------- 
\begin{figure*}
\centering
%\scalebox{0.75}{\includegraphics{Fig4Nf.eps}}  
\scalebox{0.75}{\includegraphics{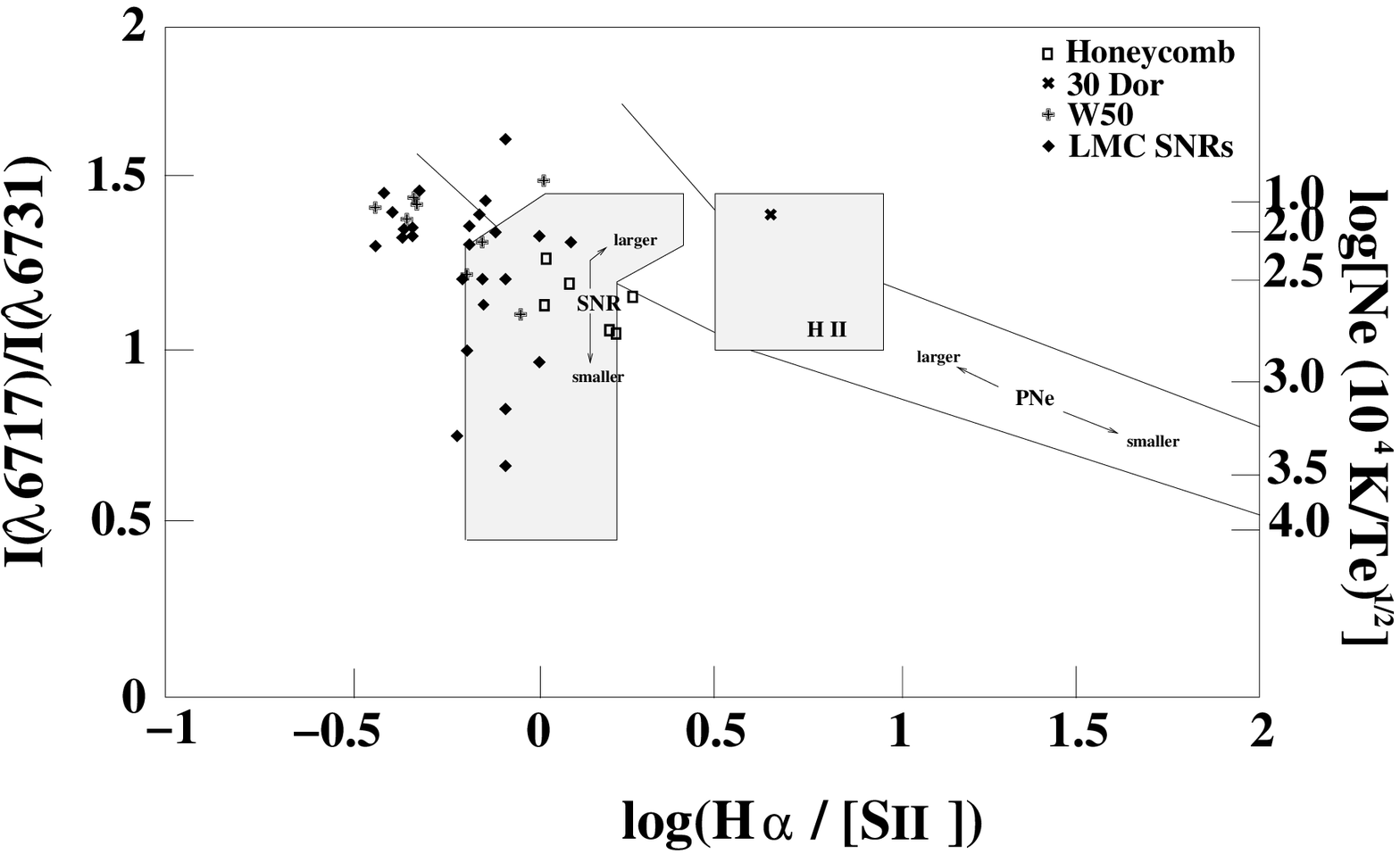}}  
\caption[]{Similar diagnostic diagram to that in Fig. 3 
but no values for the WR nebulae were available.}
\label{fig4} 
\end{figure*}
%----------------------------------------------------------- 
\begin{figure*}
\centering
%\scalebox{0.75}{\includegraphics{Fig5Nf.eps}}  
\scalebox{0.75}{\includegraphics{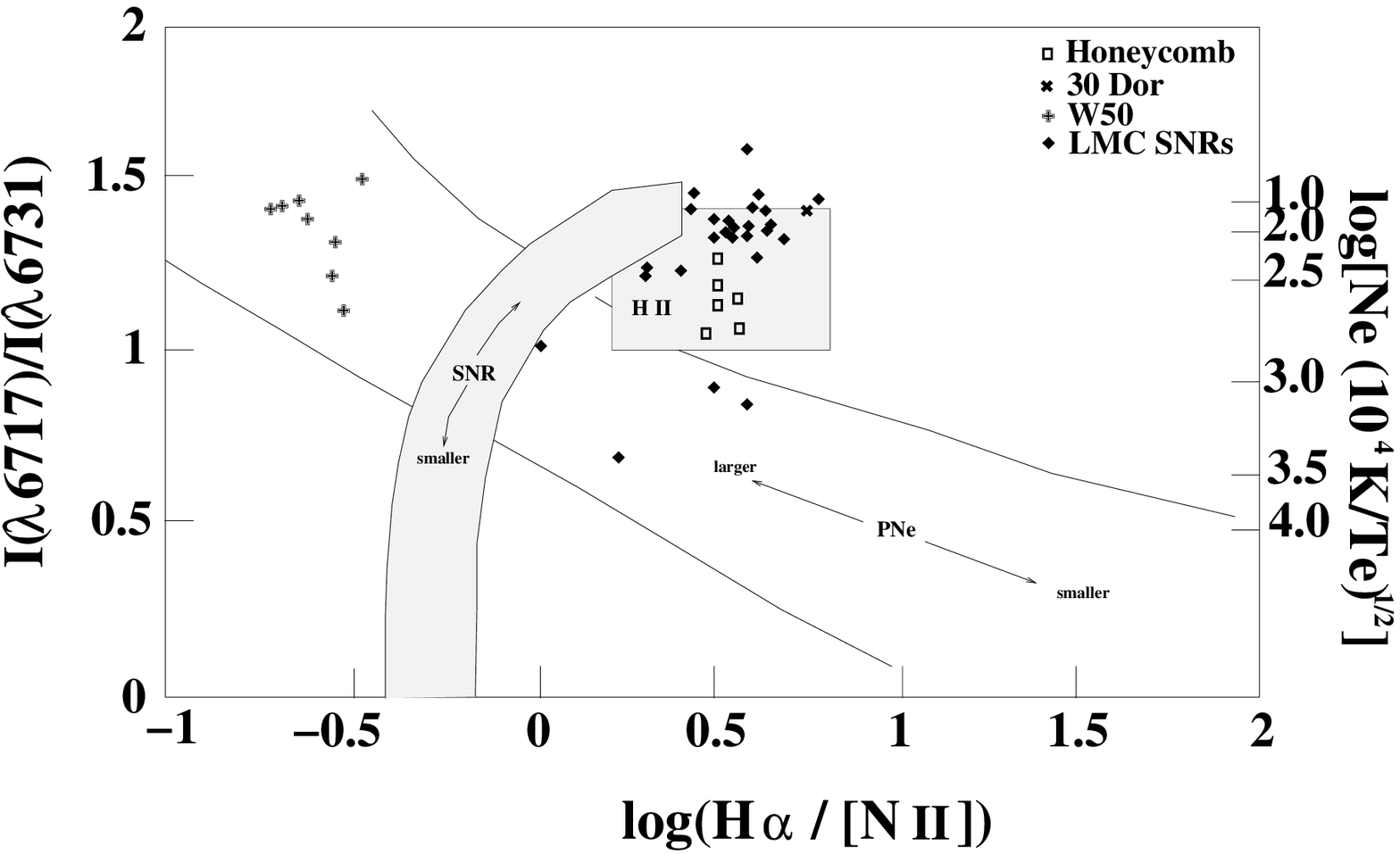}}  
\caption[]{Similar diagnostic diagram to that in Fig. 3 but again 
(as for Fig. 4) no values for the WR nebulae were available.}
\label{fig5} 
\end{figure*}
%------------------------------------------------------------
\begin{figure*}
\centering
\scalebox{0.75}{\includegraphics{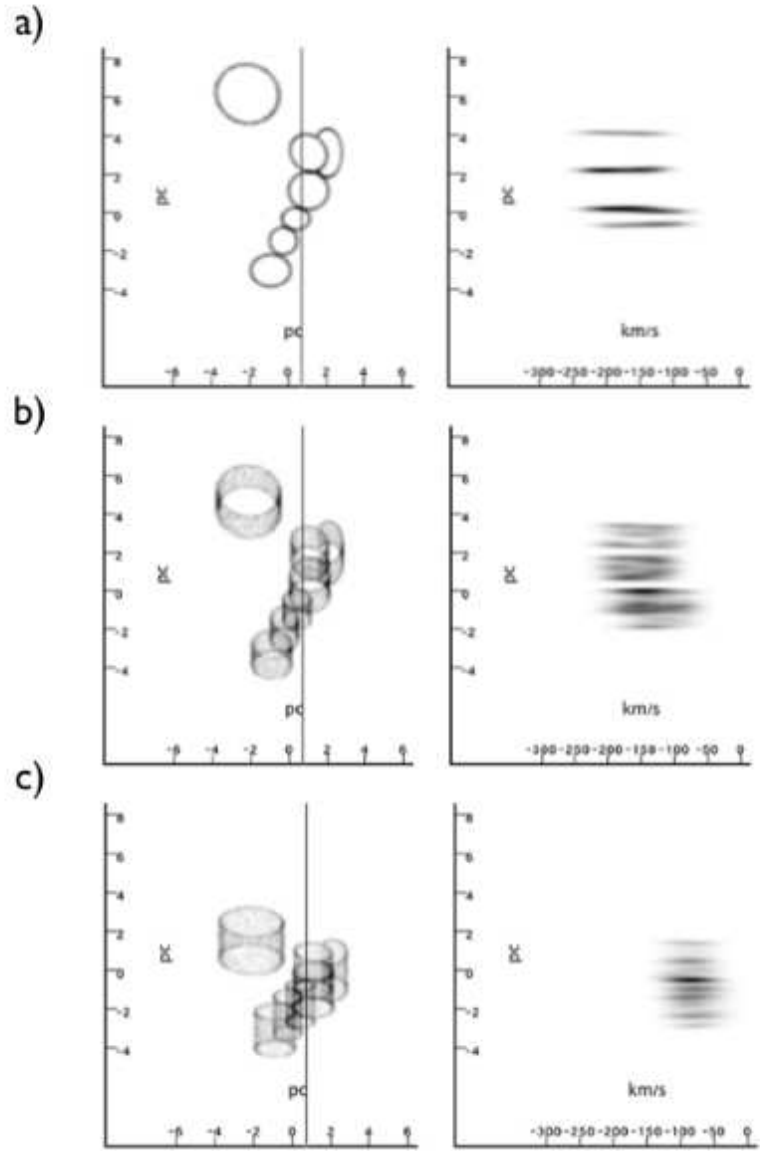}}
%\scalebox{0.75}{\includegraphics{Fig6b.ps}}\\{\scalebox{0.75}
%{\includegraphics{Fig6c.ps}}}
\caption[]{a) SHAPE code model of the Honeycomb as a set of cylinders
viewed end--on. The image is on the left and the synthetic
pv array on the right for the slit position marked in Fig. 2a. 
Sharp `velocity spikes' are generated. 
The synthetic images and pv arrays in (b) and (c) are for the 
same
model as in Figure 6a but viewed at angles of 45 and 60 degrees
respectively. The
Honeycomb structure loses coherence and the velocity spikes become
shortened and blended. The radial velocity scale is with respect to
systemic value for this region of the 30 Doradus halo.}
\label{fig6} 
\end{figure*}

\begin{figure*}
\centering
\scalebox{0.5}{\includegraphics{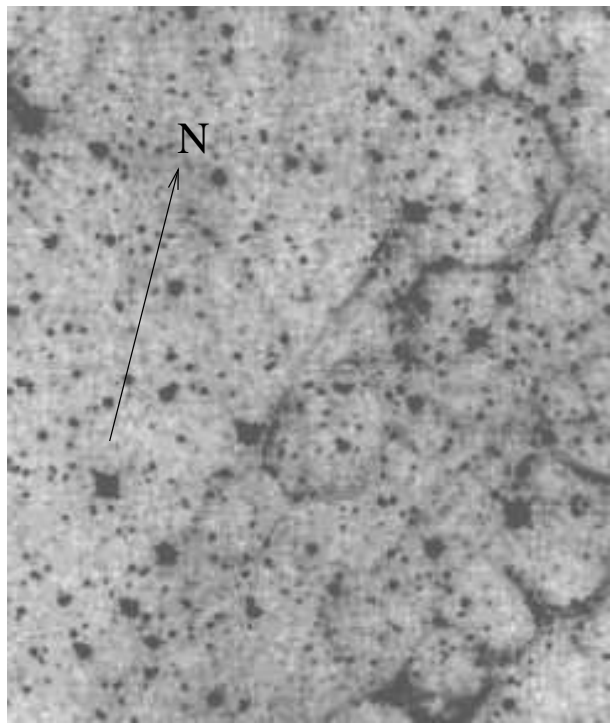}}  
\caption[]{ The broadband 
HST (675 nm) image of part of the field shown in Fig.1
a-c. The low  edge width to diameter  ratio of the most prominant Honeycomb
shell (top right) should be compared to its synthetic image in Fig. 6a.}
\label{fig7}
\end{figure*}

\begin{figure*}
\centering
\scalebox{0.75}{\includegraphics{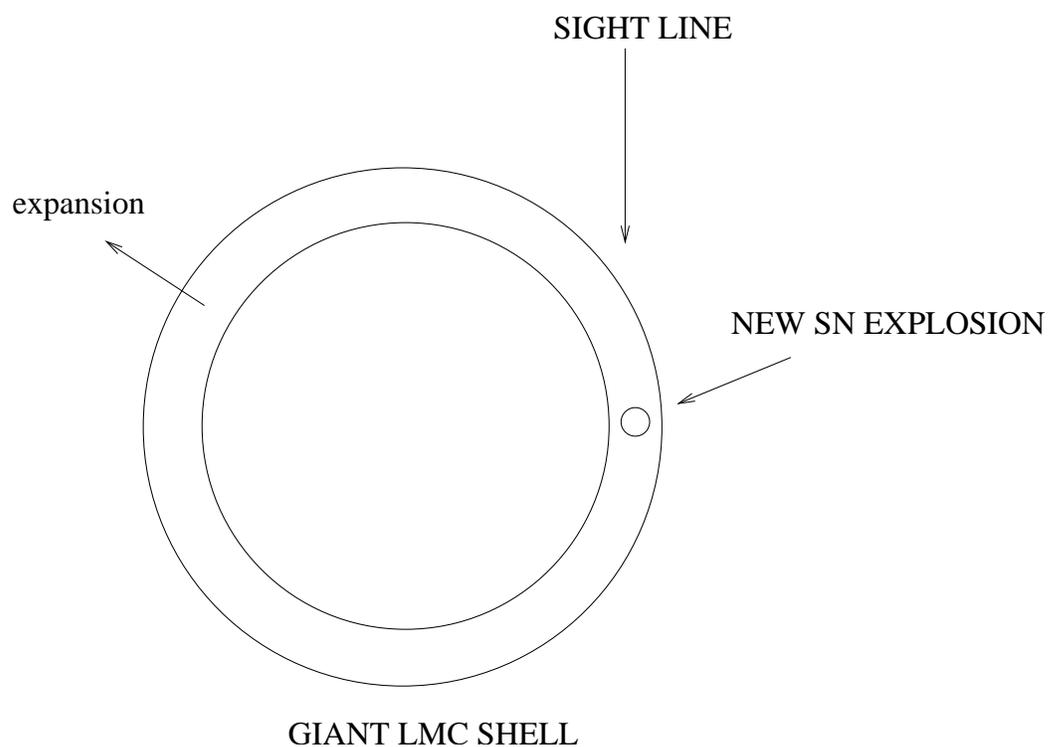}}
\caption[]{The positive and negative high--speed velocity spikes in pv
arrays can be generated by a young SNR in the edge of an LMC giant
shell. A more realistic depiction would have clumpy material in the
giant shell and it to have a more irregular structure than sketched
here. Positive and negative flows in the cylindrical cells around
clumps would be seen along the sight--line marked here.}
\label{fig8} 
\end{figure*}

\begin{figure*}
\centering
\scalebox{3}{\includegraphics{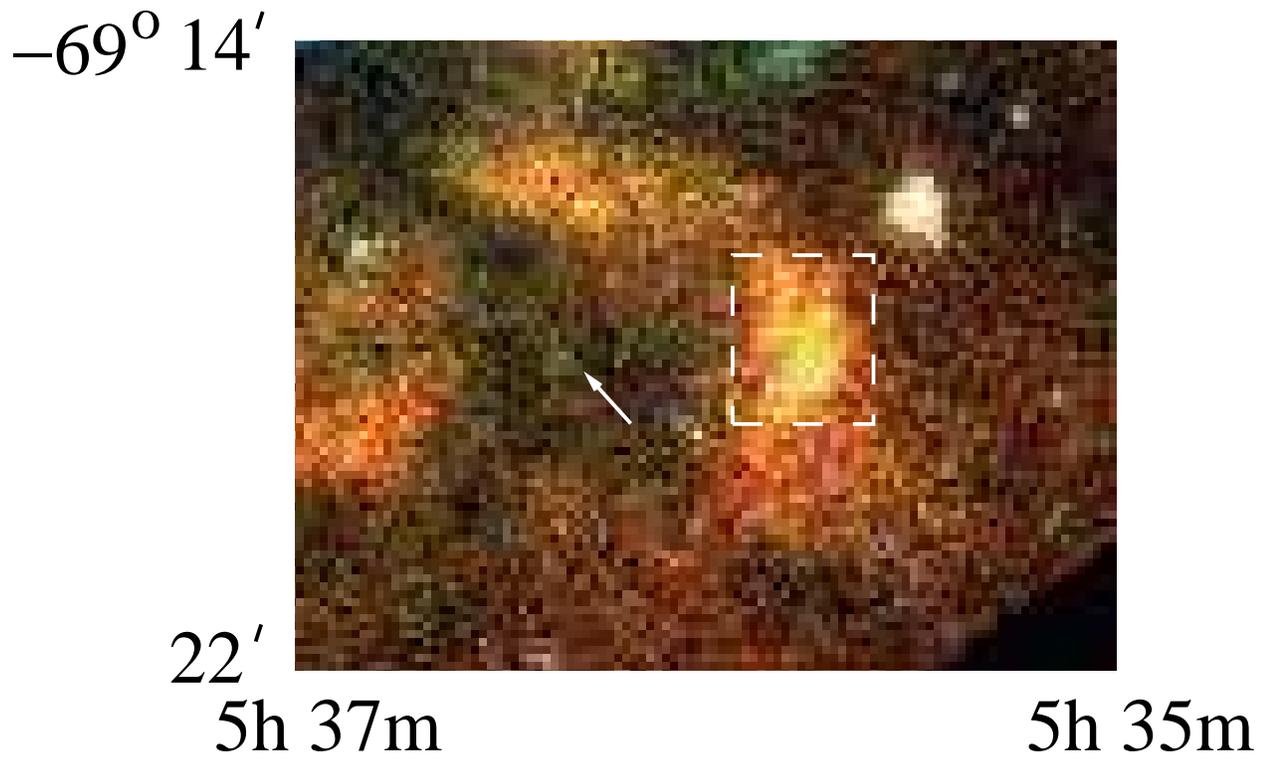}}  
\caption[]{The area of Figs.1a--c is shown as a dashed box against a
subset of the XMM X--ray image from Dennerl et al. (2001). Softer
X--rays become orange in this presentation and hard blue. SN1987A is
the bright source towards the top right and the point X--ray source
number 1 of Haberl et al. (2001) is towards the top left (J2000
coords).An arrow points to a marginally detected point source
which is more apparent in the original image of Dennerl et al. (2001)}
\label{fig9} 
\end{figure*}
%--------------------------------------------------------------
\bsp
\label{lastpage}  
\end{document}